\documentclass[preprint,12pt]{elsarticle}
\usepackage[T1]{fontenc}
\usepackage{wrapfig}
\usepackage{graphicx}
\usepackage{amssymb}
\usepackage{amsfonts}
\usepackage{amsmath}
\usepackage{fontenc}
\usepackage{longtable}
\usepackage{rotating}
\usepackage{lscape}
\usepackage{epsfig}
\usepackage{multirow}
\usepackage{hyperref}
\usepackage{lineno}
\usepackage{hyphenat}
\usepackage{subcaption}
\usepackage{cancel}
\usepackage{multirow}
\usepackage{xcolor}

\hypersetup{
    colorlinks=true,
    linkcolor=blue,
    filecolor=magenta,      
    urlcolor=cyan
    }

\newenvironment{changemargin}[2]{%
\begin{list}{}{%
\setlength{\topsep}{0pt}%
\setlength{\leftmargin}{#1}%
\setlength{\rightmargin}{#2}%
\setlength{\listparindent}{\parindent}%
\setlength{\itemindent}{\parindent}%
\setlength{\parsep}{\parskip}%
}%
\item[]}{\end{list}}

\newcommand{\CC}{C\nolinebreak\hspace{-.05em}\raisebox{.4ex}{\tiny\bf +}\nolinebreak\hspace{-.10em}\raisebox{.4ex}{\tiny\bf +}}

\newcounter{bla}

\newcommand{\generator}{Upcgen}

\journal{Computer Physics Communications}

\begin{document}

\begin{frontmatter}

\title{Upcgen: a Monte Carlo simulation program for dilepton pair production in ultra-peripheral collisions of heavy ions}

\author[pnpi]{Nazar Burmasov\corref{author}}
\author[pnpi]{Evgeny Kryshen}
\author[smi]{Paul B\"uhler}
\author[smi]{Roman Lavicka}

\cortext[author] {Corresponding author.\\\textit{E-mail address:} nazar.burmasov@cern.ch}
\address[pnpi]{Petersburg Nuclear Physics Institute named by B.P.Konstantinov of National Research Center <<Kurchatov Institute>>, 1 mkr. Orlova roshcha, 188300 Gatchina, Russia}
\address[smi]{Stefan Meyer Institute for Subatomic Physics, Kegelgasse 27, 1030 Vienna,\,Austria}

\begin{abstract}
Ultra-peripheral collisions (UPCs) of heavy ions can be used as a clean environment to study two-photon induced interactions such as dilepton pair photoproduction. Recently, precise data on lepton pair production in UPCs were obtained by the ATLAS experiment at the LHC where significant deviations, of up to 20\%, from available theoretical predictions were observed. In this work, we present a Monte Carlo event generator, \generator, that implements a refined treatment of the photon flux allowing us to improve the agreement with ATLAS data at large dilepton rapidities. Besides, the new generator offers a possibility to study photon polarization effects and set arbitrary values of the lepton anomalous magnetic moment that can be used in the future studies of tau $g-2$ via ditau production measurements in UPCs. 
\end{abstract}


\end{frontmatter}

{\bf PROGRAM SUMMARY}

\begin{small}
\noindent
{\em Program Title:}~\generator
\\
{\em CPC Library link to program files:} (to be added by Technical Editor)
\\
{\em Code Ocean capsule:} (to be added by Technical Editor)
\\
{\em Licensing provisions:} GPLv3
\\
{\em Programming language:} \CC
\\
{\em Requirements:} ROOT software toolkit; optionally: Pythia8 and (or) Pythia6 event generator, a compiler with OpenMP support.
\\
{\em Nature of problem:}
In view of deviations between theoretical predictions and new experimental measurements of dilepton production cross sections in ultra-peripheral collisions of heavy ions, a more accurate calculation is needed. Precise predictions also become crucial for the studies related to tau anomalous magnetic moment measurements via ditau pair production in UPCs. We made an attempt to implement a dedicated Monte Carlo event generator with improved calculation of the photon flux that can be used to generate dilepton pairs in UPCs with a possibility to freely change the value of the anomalous magnetic moment.
\\
{\em Solution method:}
Utilizing classes implemented in ROOT, the program calculates the dilepton pair production cross section $AA\hspace{-1mm}\to\hspace{-1mm}AA\hspace{-1mm}+\hspace{-1mm}\ell\ell$ by folding the elementary $\gamma\gamma\hspace{-1mm}\to\hspace{-1mm}\ell\ell$ cross section and the two-photon luminosity. Computation of the elementary cross section with an arbitrary value of the anomalous magnetic moment is based on the generalized vertex formalism. The calculation of the two-photon luminosity is based on an improved photon flux treatment based on realistic nuclear form factors. Since the computation of the two-photon luminosity is a time-consuming operation, a corresponding 2D-histogram is cached in a ROOT-file. The dilepton pair production cross section is then used to generate lepton pairs via a Monte Carlo simulation process. For the decay of the taus, Pythia8 or Pythia6 can be used.
\\
{\em Additional comments:}
The program is aimed on simulation of dilepton pair production in ultra-peripheral collisions of heavy ions in collider experiments. A user can set the energy of the colliding system, change species of incoming nuclei and also tune the anomalous magnetic moment of the lepton to be produced. A possibility to generate a realistic transverse momentum distribution of dilepton pairs is also taken into account.
\\
{\em References:} \href{https://github.com/nburmaso/upcgen}{https://github.com/nburmaso/upcgen} and references in this article.
\\
\end{small}

\section{Introduction}
\label{section:intro}

An important property of ultra-peripheral collisions (UPCs) of heavy ions is the large impact parameter, greater than the sum of radii of incoming nuclei~\cite{BaltzUpc,ContrerasUpc}. Under such conditions, hadronic interactions are strongly suppressed and the role of electromagnetic interactions is increased. Relativistic heavy ions are surrounded by extremely strong electromagnetic fields, which can be described in terms of the Weizs\"acker-Williams formalism and treated as a flux of quasi-real photons with very small virtuality $q^2 < (\hbar c / R_A)^2$ where $R_A$ is the radius of a nucleus. The magnitude of the photon flux is proportional to the nuclear charge squared $Z^2$, therefore the probability for photon-photon reactions scales as $Z^4$.

Considering these unique characteristics, UPCs can be used to study two-photon interactions in an effectively clean environment. One of the latest examples is the  evidence for light-by-light scattering first reported by the ATLAS and CMS collaborations at the Large Hadron Collider~(LHC)~\cite{ATLASLbyL,CMSLbyL}. 

\begin{figure}[t]
	\centering
	\includegraphics[width=0.35\textwidth]{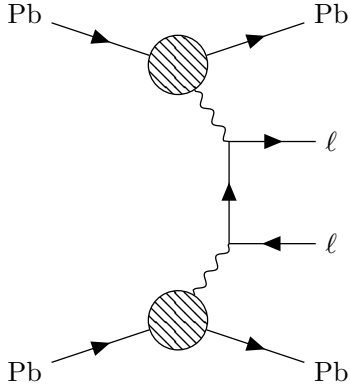}
	\caption{Production of lepton pair in ultra-peripheral collision.}
	\label{fig:lep_production}
\end{figure}

A much more abundant process 
is the production of lepton pairs in UPCs shown in Fig.~\ref{fig:lep_production}. A great amount of experimental data on photoproduction of dielectron and dimuon pairs was obtained by the ALICE, ATLAS and CMS experiments~\cite{ATLASLbyL,CMSLbyL,DimuonATLAS,ALICE_dielectrons}. Fiducial cross sections for the dimuon pair production extracted by ATLAS show significant discrepancies,  up to 20\%, with theoretical predictions obtained using the STARlight event generator~\cite{Klein:2016yzr} which is widely used for the simulation of a variety of final states in ultra-peripheral heavy-ion collisions. STARlight calculations of the photon flux from nuclei are based on several approximations therefore a more precise treatment is required especially in view of high precision data expected in the LHC Run 3 and 4~\cite{Citron:2018lsq}.

The process of dilepton production in heavy ion UPCs have been recently added in the latest version of SuperChic~\cite{Harland-Lang:2018iur,Harland-Lang:2020veo}, a Fortran based Monte Carlo event generator for central diffractive and photon-initiated production. In contrast to STARlight which was found to undershoot the ATLAS measurements by 20\% at large rapidities, the baseline SuperChic predictions overshoot the measured cross sections by about~10\%~\cite{Harland-Lang:2021ysd}.
It was shown that the dominant reason for the differences between SuperChic and STARlight results is due to an unphysical cut $b_{\gamma} > R_A$ on the impact parameter $b_{\gamma}$ of the emitted photon, that is applied in STARlight, effectively excluding the cases when dileptons are produced within one of the nuclei. Indeed, as was discussed in~\cite{Harland-Lang:2021ysd,Zha:2021jhf,Azevedo:2019fyz}, the produced lepton pair would have no hadronic and negligible electromagnetic interaction with the nuclei and would barely cause the nuclei to dissociate, therefore accounting for the cases when dileptons are produced within one of the nuclei is well justified.

Besides, it might be important to take into account photon polarization effects. Indeed, polarization vectors of the two emitted photons can be either parallel or perpendicular to each other. The two-photon luminosities for these two cases appear to have different impact-parameter dependence resulting in non-negligible differences for dilepton production cross sections compared to calculations where this effect is ignored~\cite{Baur:1990fx,Vidovic:1992ik}.

Precise treatment of photon fluxes and photon polarization effects is particularly important for the ditau production that has been proposed as a tool to measure the anomalous magnetic moment $a_\tau$ of the $\tau$ lepton~\cite{Beresford,DYNDAL2020135682,DELAGUILA1991256}. The best limits up to this day were set by the DELPHI collaboration more than 15 years ago in the studies of ditau production in $ee\to ee\tau\tau$ process~\cite{DELPHITauLimits}:

\begin{equation}
	-0.052 < a_{\tau} < 0.013 \, (95\%\,{\rm  CL}).
\end{equation}

High precision measurements of lepton magnetic moments $g_\ell = 2(1 + a_\ell)$ can be used to verify QED predictions and search for possible deviations of theoretical calculations from experimental results which may reveal the presence of effects of physics beyond the Standard Model~(BSM) such as composite nature of leptons or contributions of supersymmetric particles~\cite{compositeLeps,susyLeps}. The latest results from the Muon $g-2$ experiment at Fermilab show significant discrepancy of four standard deviations with the Standard Model predictions~\cite{FermilabGm2} revealing possible signs of BSM physics. Measurements of the anomalous magnetic moment of $\tau$ lepton might be particularly interesting in this respect, since $a_\tau$ may be up to 280 times more sensitive to BSM effects in comparison to $a_\mu$ due to  much higher mass of the $\tau$ lepton~\cite{susyLeps}.

Precise estimates of dilepton production cross sections are crucial in a variety of measurements where lepton pairs emerge as background, e.g. light-by-light and vector meson photoproduction measurements. In view of differences between data and available simulations and in view of high precision experimental data expected in Run 3 and Run 4 at the LHC, we developed a Monte Carlo event generator, \generator, dedicated to dilepton pair production in UPCs in an attempt to improve the agreement of theoretical predictions with experimental results. The calculation of nuclear $AA\hspace{-1mm}\to\hspace{-1mm}AA\hspace{-1mm}+\hspace{-1mm}\ell\ell$ cross sections is based on a refined treatment of the photon flux based on realistic nuclear form factors including cases when dileptons are produced within one of the nuclei. Photon polarization effects are also taken into account. The generator implements the generalized vertex approach for the calculation of elementary cross sections of dilepton production $\gamma\gamma\hspace{-1mm}\to\hspace{-1mm}\ell\ell$ that allows a user to choose an arbitrary value of the anomalous magnetic moment, which can be beneficial in scope of precise measurements of this parameter in LHC experiments.

In this article, we thoroughly describe the theoretical framework, that was implemented in the program; we describe the structure of the program, the simulation process and input parameters that allow a user to fine-tune calculations.

\section{Theoretical framework}
\label{section:theoretical_framework}

In the  Weizs\"acker-Williams approximation, the cross section for the dilepton production in UPCs can be obtained by convolution of the elementary $\gamma\gamma\hspace{-1mm}\to\hspace{-1mm}\ell\ell$ cross section and fluxes of photons from colliding nuclei. The calculation of the elementary cross section with an arbitrary value of the lepton anomalous magnetic moment is reviewed in Section~\ref{sec:elementary}. The calculation of the dilepton production in UPCs without accounting for photon polarization effects is described in Section~\ref{sec:nuclear}. Photon polarization effects are considered in Section~\ref{sec:polarization}. Simulation of photon transverse momentum spectra and final state radiation effects are discussed in Section~\ref{sec:pt}. Validation of the algorithms and comparison with experimental data are presented in Section~\ref{sec:validation}.

\subsection{Elementary cross section with an arbitrary $a_\ell$ value}
\label{sec:elementary}
\begin{figure}[ht]
	\centering
	\includegraphics[width=0.55\textwidth]{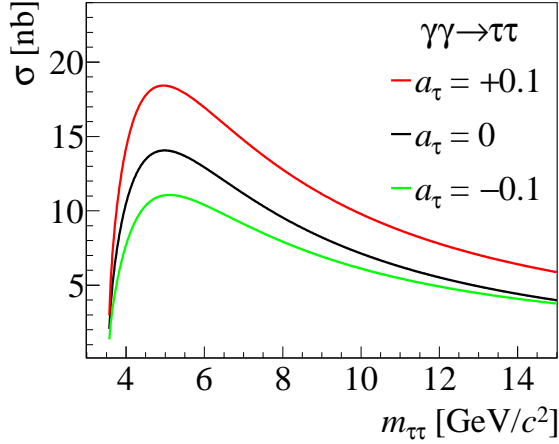}
	\caption{(Color online) Elementary $\gamma\gamma\hspace{-1mm}\to\hspace{-1mm}\tau\tau$ cross sections as functions of the ditau invariant mass for different $a_{\tau}$ values.}
	\label{fig:elem_cross_section}
\end{figure}

In order to obtain the elementary cross section of the process $\gamma\gamma \to \ell\ell$ for arbitrary $a_\ell$ values, a dedicated calculation in terms of generalized vertices was carried out following the approach proposed in~\cite{DYNDAL2020135682}. The cross section can be expressed with the formula:
\begin{equation}
	\frac{\mathrm{d}\sigma(\gamma\gamma\hspace{-1mm}\to\hspace{-1mm}\ell\ell)}{\mathrm{d} z} = \frac{2\pi}{64\pi^2s} \frac{|{\vec p}_{\ell}|}{|\vec p_{\gamma}|}
	\frac{1}{4} \sum_{\rm{spin}} \left| \mathcal{M} \right|^2 \, ,
\end{equation}
where $z = \mathrm{cos}\,\theta$ and $\theta$ is the angle of the outgoing lepton in the final state relative to the beam direction in the photon-photon center-of-mass frame, $s$ is the squared invariant mass of two photons, $p_{\ell}$ and $p_{\gamma}$ are the momenta of the lepton and photon respectively. The amplitude of the reaction $\mathcal{M}$ can be obtained using the formula~\cite{DYNDAL2020135682}:
\begin{eqnarray}
	&{\mathcal M}=(-i)\,
	\epsilon_{1 \mu}
	\epsilon_{2 \nu}
	\,\bar{u}(p_{3}) 
	\Big(
	i\Gamma^{(\gamma \ell\ell)\,\mu}(p_1)
	\frac{i(\cancel{p}_{t} + m_{\ell})}{p_t^2 - m_{\ell}^2+i\epsilon}
	i\Gamma^{(\gamma \ell\ell)\,\nu}(p_2) \nonumber\\
	&
	+
	i\Gamma^{(\gamma \ell\ell)\,\nu}(p_2)
	\frac{i(\cancel{p}_{u} + m_{\ell})}{p_u^2 - m_{\ell}^2+i\epsilon}
	i\Gamma^{(\gamma \ell\ell)\,\mu}(p_1) \Big)
	v(p_{4}) \,.
	\label{eq:amplitude}
\end{eqnarray}
Here, $p_1$ and $p_2$ are the four-momenta of the photons, $p_3$ and $p_4$ are the 4-momenta of the produced leptons, $\epsilon_{1 \mu}$ and $\epsilon_{2 \nu}$ are the polarization vectors of the photons, $p_t = p_2 - p_4$, $p_u = p_1 - p_4$, $\Gamma^{(\gamma \ell\ell)}$ is the generalized vertex function~\cite{DYNDAL2020135682}, which depends on the magnitude of the momentum transfer $q$:
\begin{equation}
	i\Gamma^{(\gamma \ell\ell)}_{\mu}(q) = 
	-ie\left[ \gamma_{\mu} F_{1}(q^{2})+ \frac{i}{2 m_{\ell}}
	\sigma_{\mu \nu} q^{\nu} F_{2}(q^{2})
	\right]
	\,,
	\label{eq:vertex}
\end{equation}
where $m_\ell$ is the mass of a lepton, $F_1(q^2)$ and $F_2(q^2)$ are the Dirac and Pauli form factors respectively. In an asymptotic limit $q^2 \to 0$ these form factors are equal to $F_1(0)=1$ and $F_2(0)=a_\ell$. This requirement is well fulfilled in UPCs, where photon virtuality is of order of $(\hbar c/R_A)^2 \sim 10^{-3}$~GeV$^2$. Then the elementary cross section $\gamma\gamma\to\ell\ell$ can be expressed as a fourth-order polynomial of $a_\ell$. At $a_\ell=0$, one can reproduce the Breit-Wheeler formula~\cite{Brodsky}:
\begin{equation}
        \frac{\mathrm{d}\sigma}{\mathrm{d}z} = \frac{\pi \alpha^2}{s}\,\beta\,\left( 2 + 4\beta^2\frac{\beta^2(1-z^2)z^2+1-\beta^2}{(1-\beta^2 z^2)^2} \right)\,,
        \qquad
        \beta^2 = 1 - \frac{4 m_\ell^2}{s}\,,
    \label{eq:breit_wheeler}
\end{equation}
where $\alpha$ is the fine structure constant, $\beta$ is the velocity of the lepton in the rest frame of the $\gamma\gamma$ system. An example of elementary cross sections calculated for different values of anomalous magnetic moment for the case of $\tau$ lepton is shown in Fig.~\ref{fig:elem_cross_section}. The obtained elementary cross sections are in good agreement with~\cite{DYNDAL2020135682}.

\subsection{Dilepton production cross section in UPC}
\label{sec:nuclear}
The cross section of the dilepton production process $AA\hspace{-1mm}\to\hspace{-1mm}AA\hspace{-1mm}+\hspace{-1mm}\ell\ell$ can be obtained by convolution of the elementary $\gamma\gamma\hspace{-1mm}\to\hspace{-1mm}\ell\ell$ cross section and the two-photon luminosity $\frac{\mathrm{d}^2 N_{\gamma\gamma}}{\mathrm{d} Y \mathrm{d} M}$:
\begin{equation}
	\frac{\mathrm{d}^2\sigma(AA\hspace{-1mm}\to\hspace{-1mm}AA\hspace{-1mm}+\hspace{-1mm}\ell\ell)}{\mathrm{d} Y \mathrm{d} M} = \frac{\mathrm{d}^2 N_{\gamma\gamma}}{\mathrm{d} Y \mathrm{d} M} \sigma(\gamma\gamma\hspace{-1mm}\to\hspace{-1mm}\ell\ell)\,,
	\label{eq:nuc_cs}
\end{equation}
where $Y$ and $M$ are the rapidity and invariant mass of the dilepton pair. Energies $k_1$ and $k_2$ of the colliding photons can be expressed in terms of $Y$ and $M$ as $k_{1,2} = \frac{M}{2}\exp(\pm Y)$.

\begin{figure}[ht]
	\centering
	\includegraphics[width=0.55\textwidth]{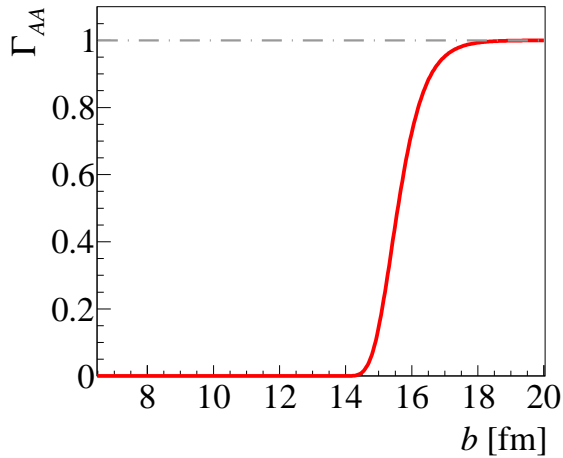}
	\caption{(Color online) The probability not to have strong interaction between incoming nuclei 
	as a function of impact parameter calculated for Pb--Pb UPC at $\sqrt{s_{\rm NN}}=5.02~{\rm TeV}$.}
	\label{fig:ggaa}
\end{figure}

The two-photon luminosity can be expressed in the photon energy phase space as
\begin{equation}
    \frac{\mathrm{d}^2 N_{\gamma\gamma}}{\mathrm{d} k_1 \mathrm{d} k_2} = \int\int \mathrm{d^2}b_{\gamma_1} \mathrm{d^2}b_{\gamma_2}\,\Gamma_{AA}(b)\,N_{\gamma A}(k_1,b_{\gamma_1})\,N_{\gamma A}(k_2,b_{\gamma_2})\,,
    \label{eq:two_phot_lumi}
\end{equation}
where $N_{\gamma A}(k_1, b_{\gamma_1})$ ($N_{\gamma A}(k_2, b_{\gamma_2})$) is the flux of photons with energy $k_1$ ($k_2$) at impact parameter $b_{\gamma_1}$ ($b_{\gamma_2}$) relative to the center of corresponding nucleus, $b = |\vec b_{\gamma_1} - \vec b_{\gamma_2}|$ is the impact parameter between two colliding nuclei, and
$\Gamma_{AA}(b)$ is the probability to not have strong interactions between the incoming nuclei at the impact parameter $b$~\cite{Guzey:2016piu}:
\begin{equation}
    \Gamma_{AA}(b)=\exp\left(-\sigma_{NN}^{\rm tot} \int \mathrm{d}^2 \Vec{b}'\,T_A(|\Vec{b}'|)\,T_A(|\Vec{b}-\Vec{b}'|) \right) \,.
    \label{eq:Gamma_AA}
\end{equation}
Here, $\sigma_{NN}^{\rm tot}$ is the total nucleon--nucleon cross section~\cite{ParticleDataGroup:2014cgo} and $T_A(b) = \int \mathrm{d}z\,\rho(b,z)$ is the optical density of a nucleus, with $\rho(b,z)$ being the nuclear density. In our calculations we use the Woods-Saxon distribution to model nuclear density. An example of the probability $\Gamma_{AA}$ calculated for Pb--Pb UPC at $\sqrt{s_{\rm NN}}=5.02~{\rm TeV}$ is shown in Fig.~\ref{fig:ggaa}. 

To calculate the two-photon luminosity, one needs to obtain photon fluxes coming from the nuclei. It can be done in a simplistic approach using a well-known expression from classical electrodynamics under assumption of point-like sources of photons~\cite{jacksonED}:
\begin{equation}
    N_{\gamma A}(k, b_\gamma) = \frac{Z^2 \alpha k^2}{\gamma^2\pi^2} \left[K_0^2(x) + \frac{1}{\gamma^2} K_1^2(x)\right] \,,
\label{eq:point_flux}
\end{equation}
where $k$ is the photon energy, $b_\gamma$ is the impact parameter of the photon relative to the nucleus center, $\gamma$ is the Lorentz boost, $x = kb_\gamma/\gamma$, and $K_0$ and $K_1$ are the modified Bessel functions. A more precise way to calculate the photon flux is based on the integral over the charge form factor of the nucleus~\cite{Vidovic:1992ik}:
\begin{equation}
    N_{\gamma A}(k,b_\gamma)=\frac{Z^2 \alpha}{\gamma \pi^2}  \left|\,\int  \frac{\mathrm{d}k_{\perp} k_{\perp}^2}{k_{\perp}^2 +
    k^2/\gamma^2}\,F_{\rm ch}(k_{\perp}^2+k^2/\gamma^2)\,J_1(b_\gamma\,k_{\perp})\,\right|^2 \,,
    \label{eq:flux_vidovic}
\end{equation}
where $F_{\rm ch}(k_{\perp}^2)$ is the charge form factor of the ion emitting the photon, 
$k_{\perp}$ is the photon transverse momentum and $J_1$ is the Bessel function of the first kind. In our calculations, we use an analytic expression for the Woods-Saxon form factor derived in~\cite{Maximon:1966sqn}:
\begin{equation}
\begin{split}
    &F_{\rm ch}(q) = \frac{4\pi^2 \rho_0 a^3}{(qa^2) \sinh^2(\pi qa)} \big[ \pi qa \cosh (\pi qa)\,\sin(qR_A)
    \\
    &- qR_A \cos(qR_A)\,\sinh (\pi qa) \big]
    + \frac{8 \pi \rho_0 a^3 \exp(-R_A/a)}{[1 + (qa)^2]^2}\,,
    \label{eq:form_factor}
\end{split}
\end{equation}
where $\rho_0$ is the nuclear density, $R_A$ is the nuclear radius,  and $a$ is the skin thickness parameter.

\begin{figure}[htb]
    \centering
    \begin{minipage}{.45\textwidth}
        \centering
        \includegraphics[width=1\linewidth]{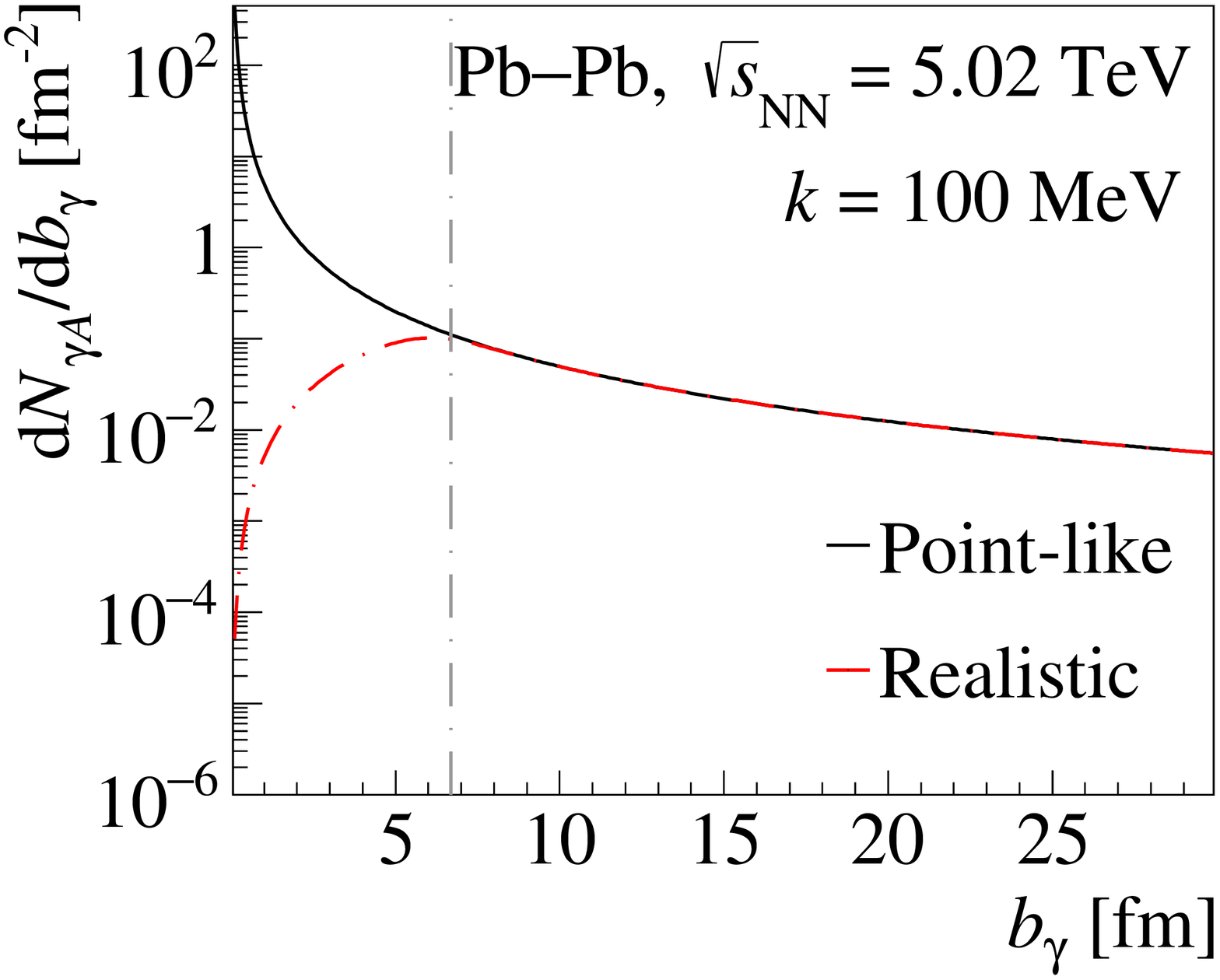}
        \\
        \vspace{-12pt}
        ~~~~\small{(a)}
    \end{minipage}
    \begin{minipage}{.45\textwidth}
        \centering
        \includegraphics[width=1\linewidth]{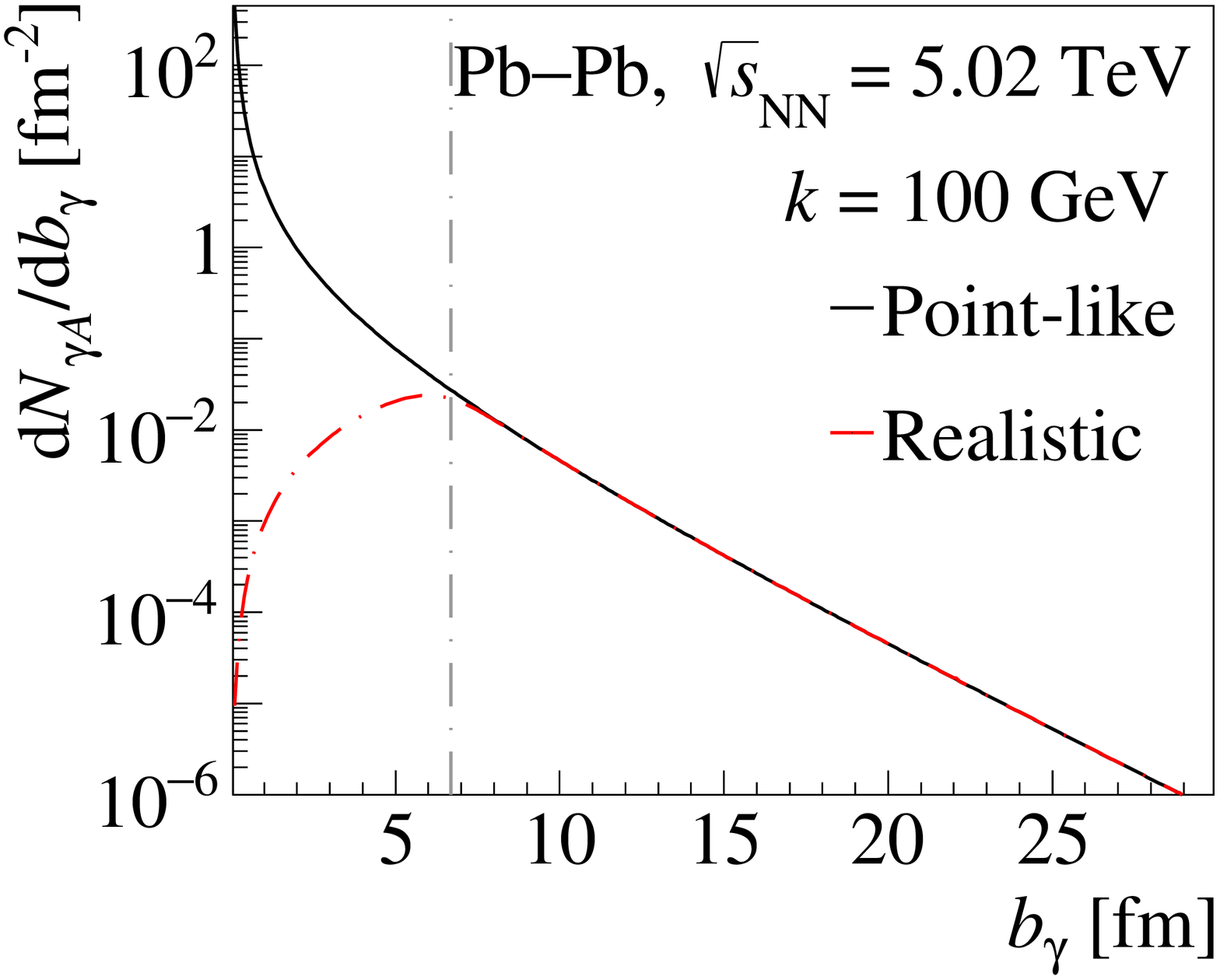}
        \\
        \vspace{-12pt}
        ~~~~\small{(b)}
    \end{minipage}
    \caption{(Color online) Photon fluxes coming from a nucleus $N_{\gamma A}$ in the point-like source approximation and the realistic description as functions of impact parameter $b_\gamma$ calculated at different photon energies: 100~MeV~(a), 100~GeV~(b).}
    \label{fig:photon_fluxes}
\end{figure}

Though at small impact parameters the magnitude of the photon flux from a point-like source strongly diverges from the realistic model (Fig.~\ref{fig:photon_fluxes}), the flux of a point-like source with additional cut-off at $b_{\gamma,\rm min} = R_{A}$ is widely used in phenomenological calculations for UPC processes, such as STARlight. This approach is well motivated in photon-nucleus interactions since the flux at impact parameters smaller than the nuclear radius is effectively suppressed by the requirement of no strong interactions between nuclei. However, in the case of photon-photon interactions, two nuclei can be separated by a large impact parameter with one of the photons still emitted at $b_{\gamma}<R_{A}$, so the cut-off at $b_{\gamma,\rm min} = R_{A}$ for the photon flux effectively rejects these cases. Therefore, as was pointed out in~\cite{Harland-Lang:2021ysd,Azevedo:2019fyz}, a realistic treatment of the form factor may become important at high photon energies. Indeed, the behavior of the photon flux as a function of the impact parameter at high photon energies becomes steeper and steeper (see comparison of photon fluxes for 100 MeV and 100 GeV photons in Fig.~\ref{fig:photon_fluxes}) and therefore flux contributions at $b_{\gamma}<R_{A}$ become more and more pronounced. See~\cite{Harland-Lang:2021ysd,Azevedo:2019fyz} for further discussion of this effect.

The two approaches of calculating the photon flux were implemented in the program allowing a user to choose between them. Evidently, the flux from a point-like source can be calculated faster, but it can lead to inaccurate results, thus in our program the default way to calculate photon fluxes is via integration of the realistic form factor.

\begin{figure}[ht]
	\centering
	\includegraphics[width=0.5\textwidth]{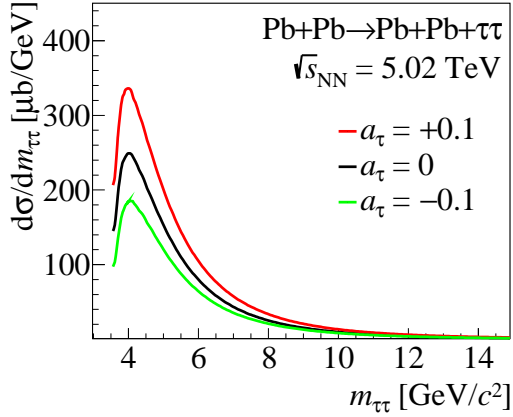}
	\caption{(Color online) Cross section for the process 
	$\mathrm{Pb\hspace{-1mm}+\hspace{-1mm}Pb}\hspace{-1mm}\to\hspace{-1mm}\mathrm{Pb\hspace{-1mm}+\hspace{-1mm}Pb}\hspace{-1mm}+\tau\tau$
	as a function of $m_{\tau\tau}$ for different $a_\tau$ values, calculated for Pb--Pb UPCs at $\sqrt{s_{\rm NN}}=5.02~{\rm TeV}$.}
	\label{fig:nuc_cross_section}
\end{figure}

With the two-photon luminosity obtained as discussed above, the calculation of the cross section for the process $AA\hspace{-1mm}\to\hspace{-1mm}AA+\ell\ell$ becomes straightforward. An example of the resulting cross sections as a function of the ditau invariant mass for different values of the anomalous magnetic moment is shown in Fig.~\ref{fig:nuc_cross_section}. As before, for this test calculation we considered ditau production in Pb--Pb UPC at $\sqrt{s_{\rm NN}}=5.02~{\rm TeV}$. The total cross section $\sigma({\rm PbPb}\hspace{-1mm}\to\hspace{-1mm}{\rm PbPb}\hspace{-1mm}+\hspace{-1mm}\tau\tau, a_\tau=0)=1.06$ mb agrees well with calculations by Dyndal \textit{et al.}~\cite{DYNDAL2020135682}.
\subsection{Photon polarization effects}
\label{sec:polarization}
Up to now, we ignored photon polarization effects. In a more detailed picture, polarization vectors of two colliding photons can be either parallel or perpendicular to each other corresponding to even-parity ($\sigma_{||}$) and odd-parity ($\sigma_{\perp}$) components of the elementary $\gamma\gamma\hspace{-1mm}\to\hspace{-1mm}\ell\ell$ cross section. Photon impact parameters $\vec b_{\gamma_1}$ and $\vec b_{\gamma_2}$ form an angle $\phi$ in the transverse plane, therefore intensities of the parallel and perpendicular cases are proportional to \break $\cos^2\phi = |\vec b_{\gamma_1} \cdot \vec b_{\gamma_2}|^2/(b_{\gamma_1} b_{\gamma_2})^2$ and $\sin^2\phi = |\vec b_{\gamma_1} \times \vec b_{\gamma_2}|^2/(b_{\gamma_1} b_{\gamma_2})^2$, respectively~\cite{Baur:1990fx,Vidovic:1992ik}.

As was shown in~\cite{Baur:1990fx,Vidovic:1992ik}, the two-photon luminosities for parallel and perpendicular polarization cases have different dependence on the impact parameter $b$ between colliding nuclei. Exclusion of small impact parameters due to veto on hadronic interactions results in significant differences between two-photon luminosities in these two cases especially at high photon energies where small $b$ become dominant. Therefore even-parity and odd-parity components of the elementary $\gamma\gamma\hspace{-1mm}\to\hspace{-1mm}\ell\ell$ cross section in general contribute to the UPC cross section with different weights.

\begin{figure}[htb]
    \centering
    \begin{changemargin}{-1cm}{0cm}
    \begin{minipage}{.5\textwidth}
        \centering
        \includegraphics[width=1.1\linewidth]{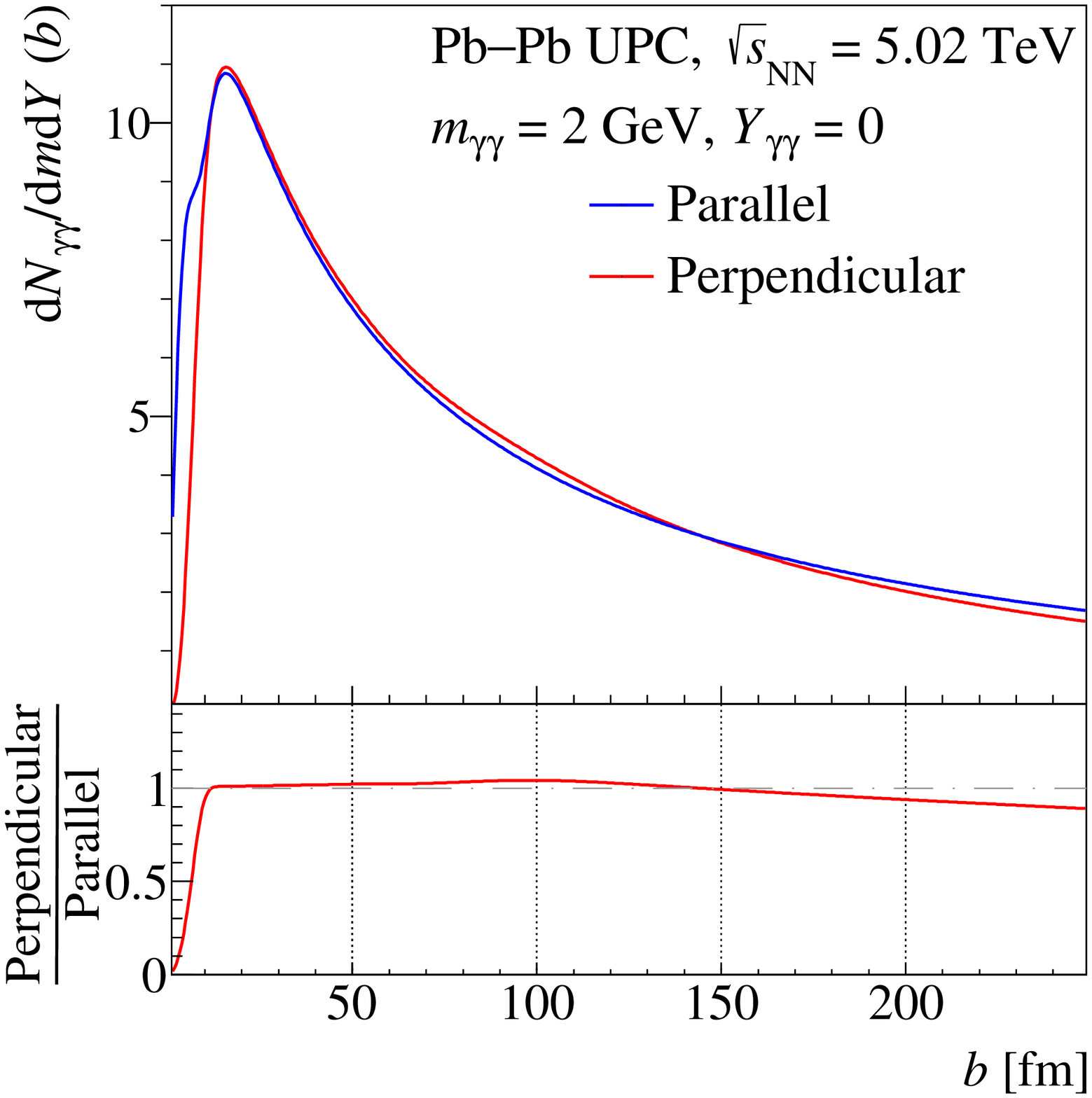}
        \\
        \vspace{-10pt}
        \hspace{30pt}\small{(a)}
    \end{minipage}%
    \hspace{16pt}
    \begin{minipage}{.5\textwidth}
        \centering
        \includegraphics[width=1.1\linewidth]{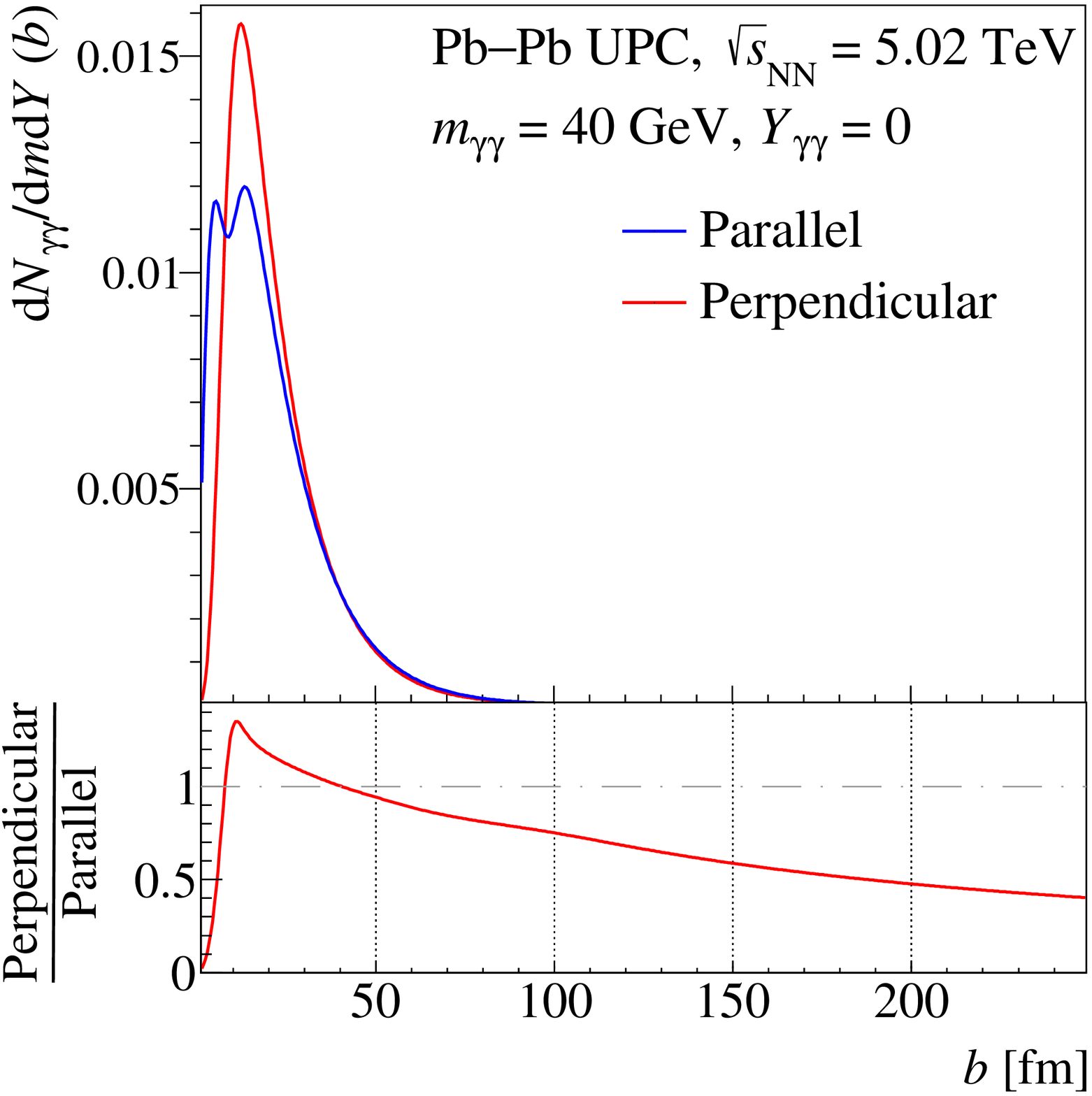}
        \\
        \vspace{-10pt}
        \hspace{30pt}\small{(b)}
    \end{minipage}
    \end{changemargin}
    \caption{(Color online) Two-photon luminosities at invariant masses $m_{\gamma\gamma} = 2$\,GeV and $m_{\gamma\gamma} = 40$\,GeV at midrapidity in Pb--Pb collisions at $\sqrt{s_{\rm NN}} = 5.02$~TeV as a function of impact parameter between colliding nuclei for parallel (blue) and perpendicular (red) photon polarization vectors.}
    \label{fig:lumi_polarized_vs_b}
\end{figure}

The cross section for the dilepton production cross section in UPCs can be decomposed into the sum of two terms:
\begin{equation}
	\frac{\mathrm{d}^2\sigma(AA\hspace{-1mm}\to\hspace{-1mm}AA\hspace{-1mm}+\hspace{-1mm}\ell\ell)}{\mathrm{d} Y \mathrm{d} M} = \frac{\mathrm{d}^2 N^{||}_{\gamma\gamma}}{\mathrm{d} Y \mathrm{d} M} \sigma_{||}(\gamma\gamma\hspace{-1mm}\to\hspace{-1mm}\ell\ell) +
	\frac{\mathrm{d}^2 N^{\perp}_{\gamma\gamma}}{\mathrm{d} Y \mathrm{d} M} \sigma_{\perp}(\gamma\gamma\hspace{-1mm}\to\hspace{-1mm}\ell\ell)\,,
	\label{eq:nuc_cs_polarized}
\end{equation}
where two-photon luminosities for parallel and perpendicular photon polarization vectors read~\cite{Baur:1990fx,Vidovic:1992ik}:
\begin{eqnarray}
    \frac{\mathrm{d}^2 N^{||}_{\gamma\gamma}}{\mathrm{d} k_1 \mathrm{d} k_2} &=& \int\int \mathrm{d^2}b_{\gamma_1} \mathrm{d^2}b_{\gamma_2}\,\Gamma_{AA}(b)\,N_{\gamma A}(k_1,b_{\gamma_1})\,N_{\gamma A}(k_2,b_{\gamma_2})
    \frac{|\vec b_{\gamma_1} \cdot \vec b_{\gamma_2}|^2}{b_{\gamma_1}^2 b_{\gamma_2}^2}\,,\quad\\
    \frac{\mathrm{d}^2 N^{\perp}_{\gamma\gamma}}{\mathrm{d} k_1 \mathrm{d} k_2} &=& \int\int \mathrm{d^2}b_{\gamma_1} \mathrm{d^2}b_{\gamma_2}\,\Gamma_{AA}(b)\,N_{\gamma A}(k_1,b_{\gamma_1})\,N_{\gamma A}(k_2,b_{\gamma_2}) 
    \frac{|\vec b_{\gamma_1} \times \vec b_{\gamma_2}|^2}{b_{\gamma_1}^2 b_{\gamma_2}^2}\,.\qquad
    \label{eq:two_phot_lumi_polarized}
\end{eqnarray}
Typical impact-parameter dependence of these two-photon luminosities is illustrated in Fig.~\ref{fig:lumi_polarized_vs_b} for photon energies $k_1=k_2=1$~GeV and $k_1=k_2=20$~GeV corresponding to $m_{\ell\ell} = 2$~GeV/$c^2$ and $m_{\ell\ell} = 40$~GeV/$c^2$, respectively, at midrapidity. The two-photon luminosity with parallel photon polarization vectors is dominant at $b\lesssim R_A$ and at large $b$ while the contribution with perpendicular polarization vectors is more pronounced at intermediate impact parameters. Accounting for $\Gamma_{AA}$ factor effectively suppresses the region $b<2R_A$ therefore the differences between the integrated luminosities with parallel and perpendicular configurations are expected to be small.

Note that if the veto on hadronic interactions is neglected, $\Gamma_{AA}(b)\equiv1$, both  two-photon luminosities become equal to one half of the total two-photon flux, so one can simply replace the sum of $\sigma_{||}$ and $\sigma_{\perp}$ with the unpolarized elementary cross section $\sigma = \frac12(\sigma_{||}+\sigma_{\perp})$ and recover the usual formula~(\ref{eq:nuc_cs}).

Even-parity and odd-parity elementary cross sections can be obtained similarly to the approach described in Section~\ref{sec:elementary}. Angular-differential $\sigma_{||}$ and $\sigma_{\perp}$ cross sections for $a_\tau=0$ read:
\begin{equation}
        \frac{\mathrm{d}\sigma_{||}}{\mathrm{d}z} = 
        \frac{2 \pi \alpha^2}{s}\,\beta\,
        \frac{ 2(2 - z^2)\beta^2 - (3-2z^2+z^4)\beta^4 }{(1-z^2\beta^2)^2}\,,
    \label{eq:cs_scalar}
\end{equation}
\begin{equation}
        \frac{\mathrm{d}\sigma_{\perp}}{\mathrm{d}z} = 
        \frac{2 \pi \alpha^2}{s}\,\beta\,
        \frac{ 2 - z^2\beta^2 + (1-z^2)^2\beta^4 }{(1-z^2\beta^2)^2}\,.
    \label{eq:cs_pesudoscalar}
\end{equation}

\begin{figure}[htb]
    \centering
    \begin{changemargin}{-1cm}{1cm}
    \begin{minipage}{.5\textwidth}
        \centering
        \includegraphics[width=1.1\linewidth]{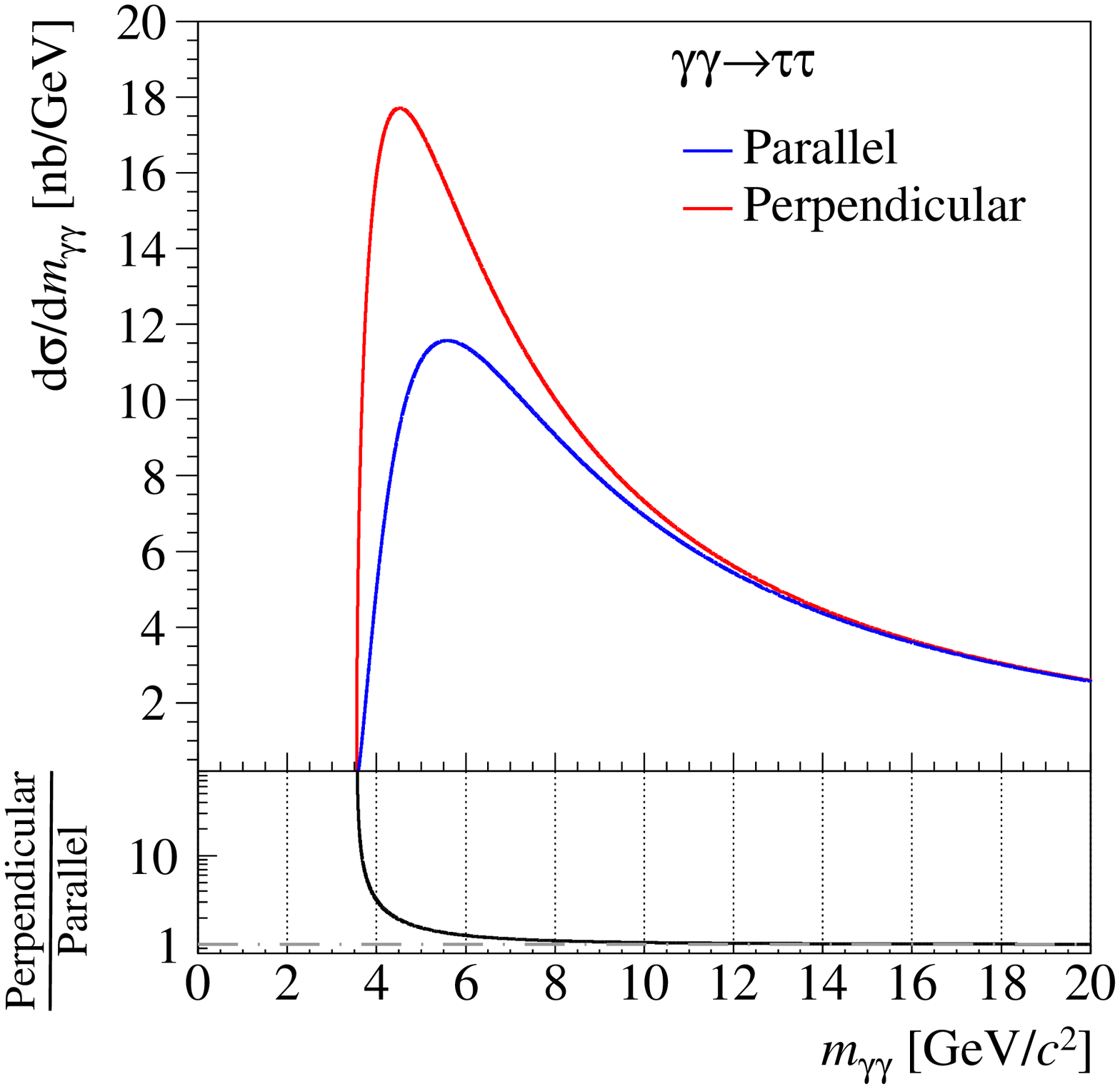}
        \\
        \vspace{-10pt}
        \hspace{30pt}\small{(a)}
    \end{minipage}%
    \hspace{16pt}
    \begin{minipage}{.5\textwidth}
        \centering
        \includegraphics[width=1.1\linewidth]{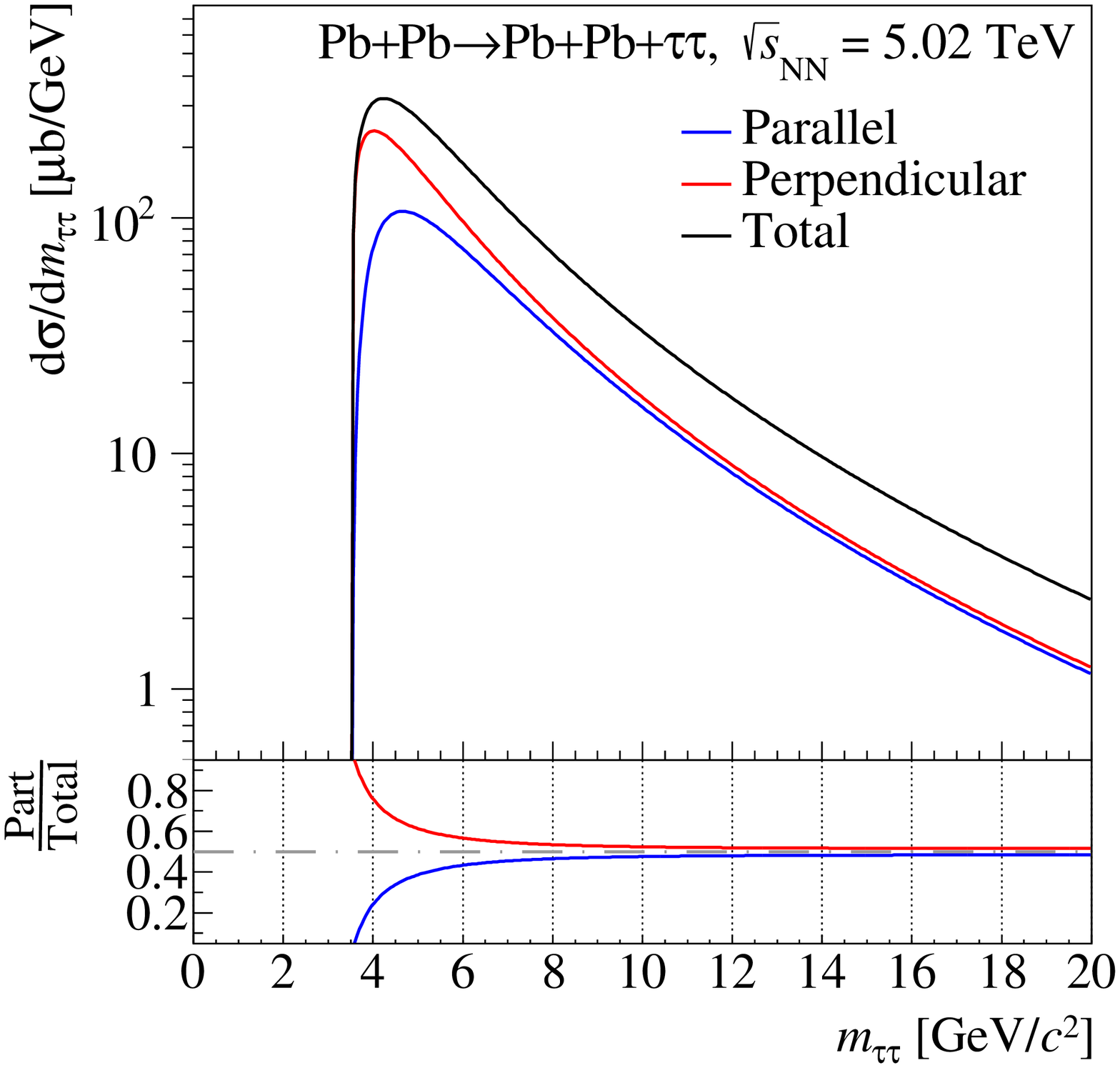}
        \\
        \vspace{-10pt}
        \hspace{30pt}\small{(b)}
    \end{minipage}
    \end{changemargin}
    \caption{(Color online) Elementary $\gamma\gamma \to \tau \tau$ cross sections (left) and nuclear ${\rm PbPb}\hspace{-1mm}\to\hspace{-1mm}{\rm PbPb}\hspace{-1mm}+\hspace{-1mm}\tau\tau$ cross sections (right) for parallel and perpendicular orientations of photon polarization vectors.}
    \label{fig:cs_scal_vs_pseudosc}
\end{figure}

\begin{figure}[htb]
\centering
\includegraphics[width=0.65\linewidth]{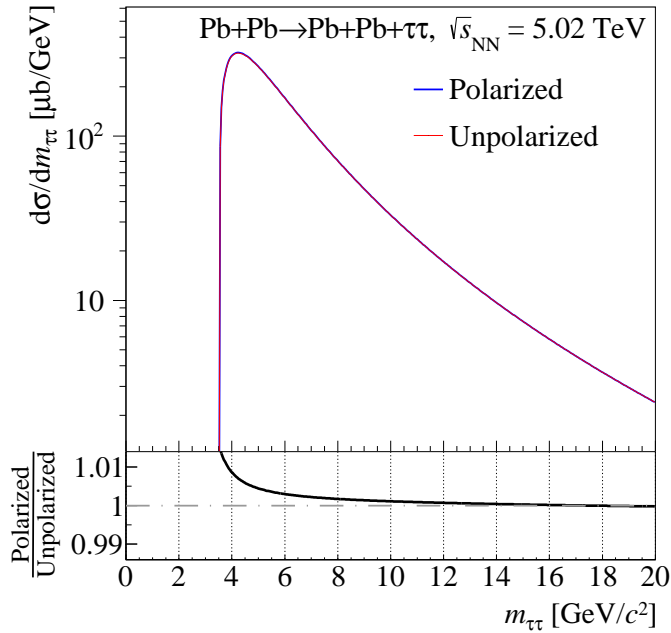}
\caption{(Color online) Comparison of tau pair production cross sections calculated with and without accounting for photon polarization effects.}
\label{fig:cs_pol_vs_unpol}
\end{figure}

Angular-integrated even-parity and odd-parity cross sections for \break $\gamma\gamma \to \tau \tau$ are shown in Fig.~\ref{fig:cs_scal_vs_pseudosc}, left. Odd-parity elementary cross section ($\sigma_{\perp}$) is dominant near threshold~\cite{Vidovic:1992ik} resulting in larger contribution of the odd-parity component in the nuclear cross section at low masses, see Fig.~\ref{fig:cs_scal_vs_pseudosc}, right. 
In the ultra-relativistic limit ($\beta \to 1$) both contributions become the same, therefore photon polarization effects have negligible impact (below per-mille level) on the calculation of dielectron and dimuon production cross sections. However, in the tau pair production, photon polarization effects result in up to 1\% differences compared to calculations in which polarization is ignored,  see Fig.~\ref{fig:cs_pol_vs_unpol}.

It is worth noting, that the photon polarization effects are also taken into account in the SuperChic generator~\cite{Harland-Lang:2018iur, Harland-Lang:2020veo}. In contrast to our approach, SuperChic calculations are performed in terms of helicity amplitudes, however the net effect is expected to be the same.

\subsection{Photon transverse momentum and final state radiation effects}
\label{sec:pt}

The transverse momentum of the final state dilepton is simulated as the vector sum of the transverse momenta $k_{1\perp}$ and $k_{2\perp}$ of the two photons. Similar to the STARlight approach~\cite{Klein:2016yzr}, $k_{\perp}$ of the photons is sampled according to the following distribution~\cite{Vidovic:1992ik}:
\begin{equation}
    \frac{d N_{\gamma A}}{d k_{\perp}} \sim  \frac{F_{\rm ch}^2(k_{\perp}^2+k^2/\gamma^2)}{(k_{\perp}^2 + k^2/\gamma^2)^2} k_{\perp}^3\,.
    \label{eq:flux_vidovic_pt}
\end{equation}

Final state radiation (FSR) from outgoing charged leptons  may result in significant broadening of dilepton $p_{\rm T}$ distributions~\cite{DimuonATLAS,Klein:2020jom}. Inspired by the ATLAS approach~\cite{DimuonATLAS}, we introduced a possibility to inject dilepton events into the built-in Pythia~8 generator with activated QED showering process and the hard scale set to one lepton’s~$p_{\rm T}$.

\subsection{Comparison with ATLAS results}
\label{sec:validation}
\begin{figure}[htb]
    \centering
    \begin{changemargin}{-1cm}{0cm}
    \begin{minipage}{.5\textwidth}
        \centering
        \includegraphics[width=1.1\linewidth]{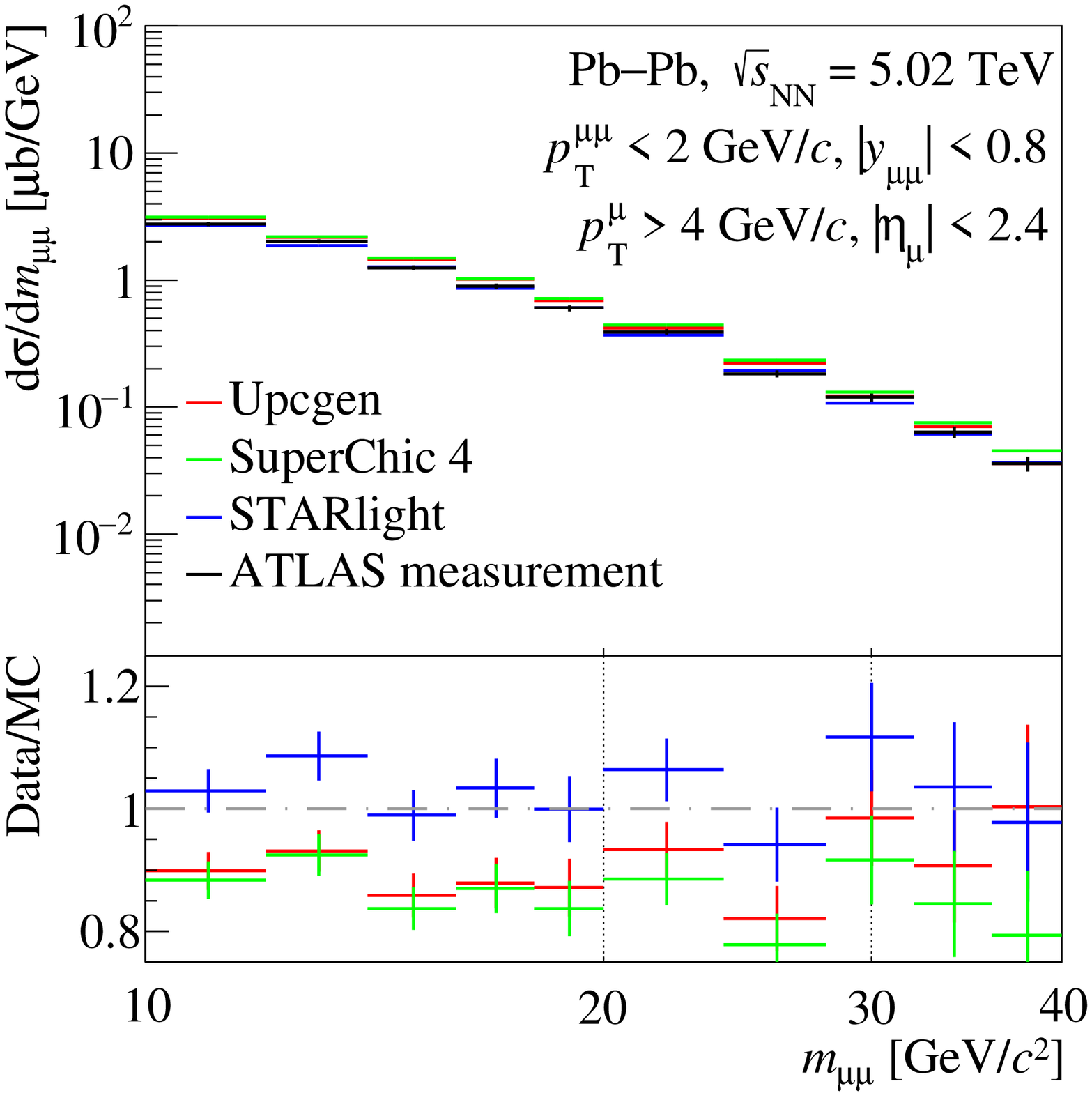}
        \\
        \vspace{-4pt}
        \hspace{30pt}\small{(a)}
    \end{minipage}%
    \hspace{16pt}
    \begin{minipage}{.5\textwidth}
        \centering
        \includegraphics[width=1.1\linewidth]{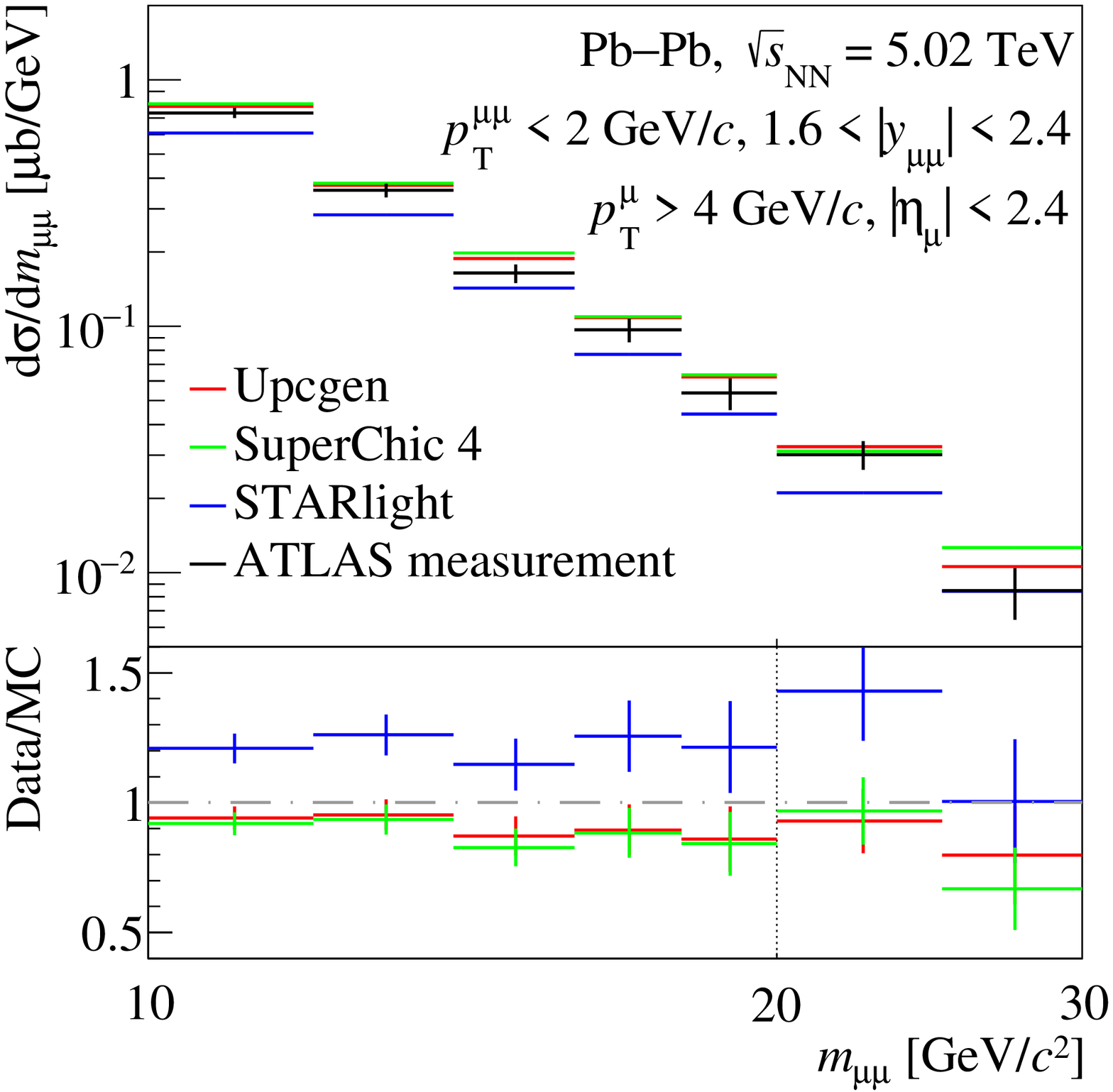}
        \\
        \vspace{-4pt}
        \hspace{30pt}\small{(b)}
    \end{minipage}
    \end{changemargin}
    \caption{(Color online) Dimuon measurements by ATLAS in comparison to estimates from \generator, SuperChic~4 and STARlight for rapidities in $|y_{\mu\mu}| < 0.8$~(a) and \break $1.6 < |y_{\mu\mu}| < 2.4$~(b).}
    \label{fig:comparison_w_data}
\end{figure}

To validate the calculations, we simulated dimuon pair photoproduction in Pb--Pb UPCs at $\sqrt{s_{\rm NN}}=5.02~{\rm TeV}$ using \generator~and applied fiducial selections on both muon pairs and single muons used by the ATLAS collaboration in the real measurement~\cite{DimuonATLAS}: $p^{\mu\mu}_{\rm T} < 2~{\rm GeV}/c$, $m_{\mu\mu} > 10~{\rm GeV}/c^2$, $p^{\mu}_{\rm T} > 4~{\rm GeV}/c$, $|\eta_{\mu}| < 2.4$. We compare our results with SuperChic calculations in the same fiducial phase space and, in addition, we reproduced calculations that ATLAS carried out with STARlight~2 using the latest version of the program, STARlight~3, in order to compare our estimates with the contemporary and commonly used model. Figure~\ref{fig:comparison_w_data} represents a comparison between the experimental results and calculations based on simulations from \generator, STARlight and SuperChic generators. We consider two regions of dimuon rapidity $y_{\mu\mu}$: central ($|y_{\mu\mu}| < 0.8$) and intermediate ($1.6 < |y_{\mu\mu}| < 2.4$). First, one can conclude that Upcgen and SuperChic generators nicely agree with each other in spite of many differences in the internal implementation. 
At central rapidities, STARlight is in good agreement with the experimental data while Upcgen and SuperChic generators overestimate the cross section by about 10\% though the difference doesn't exceed 2 standard deviations. On the other hand, at relatively high rapidities corresponding to relatively high energy of one of the photons, STARlight tends to underestimate the cross section by about 20\%, while \generator~and SuperChic stay in general closer to the data, pointing to significant contribution of photons emitted at small impact parameters $b<R_A$. See~\cite{Harland-Lang:2021ysd} for more details.

\begin{figure}[htb]
\centering
\includegraphics[width=0.7\linewidth]{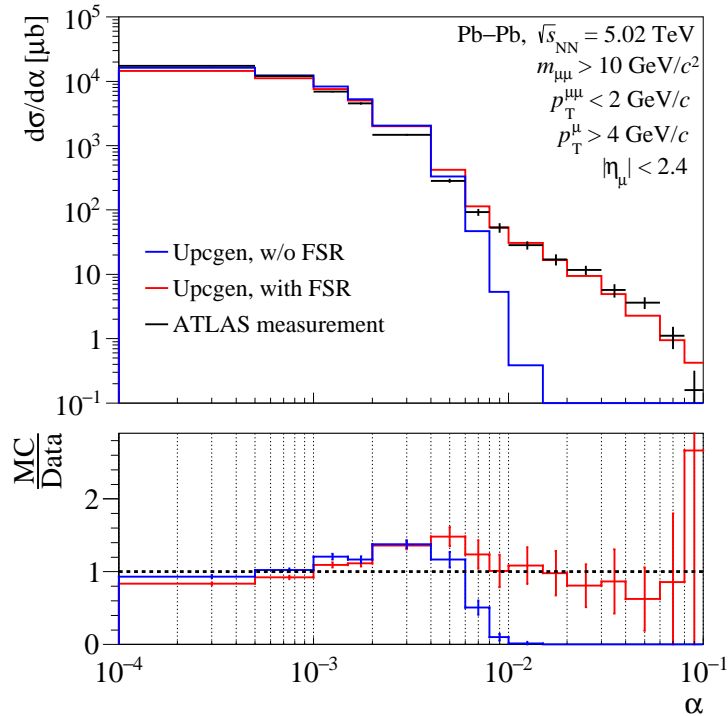}
\caption{(Color online) Dimuon acoplanarity distribution measured by ATLAS~\cite{DimuonATLAS} in Pb--Pb collisions and~\generator\ results with and without final state radiation effects.}
\label{fig:aco}
\end{figure}

The effects of non-zero dilepton $p_{\rm T}$ and FSR from outgoing leptons were checked using distribution of acoplanarity $\alpha = 1-|\Delta \varphi|/\pi$, where $\Delta \varphi$ is the difference between azimuthal angles of two leptons. The acoplanarity distribution measured by ATLAS in Pb--Pb collisions after applying the $0n0n$ condition~\cite{DimuonATLAS}, effectively vetoing additional photon exchanges between colliding nuclei, is shown in Fig.~\ref{fig:aco} in comparison with Upcgen results with and without final state radiation effects. While the option without FSR fails to describe the acoplanarity tail at $\alpha > 0.01$, the option including FSR effects  reproduces the measurement with reasonable accuracy. Still, one can notice up to 20\% discrepancies between data and \generator\,at small $\alpha < 0.01$. Similar discrepancy was obtained with the STARlight generator coupled with Pythia~8 to simulate FSR effects~\cite{DimuonATLAS}. The reason for these discrepancies can be related to the fact that \generator\ and STARlight rely on the effective photon $k_{\rm T}$ distribution averaged over all impact parameters (see~Eq.~\ref{eq:flux_vidovic_pt}). In the considered $0n0n$ UPC sample, small impact parameters are effectively suppressed therefore photon transverse momentum distributions are expected to be narrower than average. This description can be further improved in the Generalized Equivalent Photon Approximation based on Wigner distributions which contain both the transverse momentum and impact parameter information of the incoming photons~\cite{Klein:2020jom}.

This concludes an overview of the theoretical framework that was used to implement all the necessary calculations of cross sections of dilepton production in ultra-peripheral collisions of heavy ions.

\section{Program overview}
\label{section:program_overview}

In this section, an overview of the calculation process is given, as well as a detailed description of the program code.

\subsection{Simulation process}
\label{subsection:simulation_process}

In the simulation process, all necessary distributions are obtained following the order, in which the theoretical framework is presented in Section~\ref{section:theoretical_framework}, then events are generated:
\begin{itemize}
    \item the elementary $\gamma\gamma\hspace{-1mm}\to\hspace{-1mm}\ell\ell$ cross section is calculated,
    \item the two-photon luminosity is obtained and cached into a file for further usage,
    \item the nuclear cross section is calculated using the elementary cross section and the two-photon luminosity,
    \item unweighted events are generated according to the obtained nuclear cross section,
    \item optionally, decays of final state $\tau$ leptons are simulated for the case of ditau pair production.
\end{itemize}
The essential model parameters are provided by the user as program arguments. All calculations are carried out with use of classes implemented in the ROOT software toolkit~\cite{BRUN199781,root_soft}. The resulting cross sections, as well as two-photon luminosity, are stored into one-dimensional and two-dimensional histograms. The width of the bins of all axes can be chosen by the user such that the accuracy of the calculations can be tuned in order to achieve desired balance between precision and computation time.

\begin{figure}[ht]
	\centering
	\includegraphics[width=0.5\textwidth]{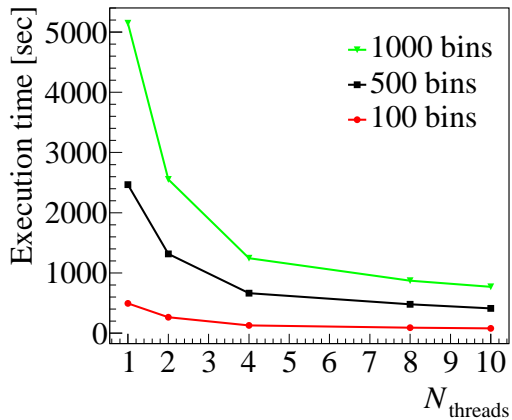}
	\caption{(Color online) Total execution time in dependence of binning along the $M$ axis and number of worker threads.}
	\label{fig:exec_times}
\end{figure}

In order to speed up the simulation process, the two-photon luminosity in different $M$ and $Y$ intervals can be calculated in parallel. An implementation of the parallel computation is simplistic and based on OpenMP technology~\cite{openmp15}: dilepton mass intervals $M$ are evenly distributed among all worker threads which calculate the luminosity values for each $M$ and $Y$ independently and store them in 2D histograms. After all calculations have finished, the thread-private histograms are merged, and the total histogram is used to calculate the nuclear cross section. To illustrate the effect of parallel computation on the total execution time, we tested the program in a typical problem of simulation of 100\,000 ditau pairs in UPCs at $\sqrt{s_{\rm NN}}=5.02~{\rm TeV}$ with masses between $\sim$4 and 50~GeV/$c^2$ for different binnings along the mass axis and different level of parallelism: 100, 500 and 1000 bins for 1, 2, 4, 8 and 10 worker threads. In this test we omitted simulation of final state lepton decays. The simulations were carried out on a computer with the 6-core Intel~i7-10750H CPU in the system ran by Ubuntu~20.04~LTS. Resulting execution times are represented in Fig.~\ref{fig:exec_times}.
Parallel computation of two-photon luminosity can help to significantly decrease the program execution time, since it may take up to 99\% of the total simulation time (omitting time needed to calculate particle decays). It is especially visible for very fine binnings, which are often required to adequately calculate cross sections that drop dramatically with decreasing dilepton mass.

In the main simulation process, a dilepton pair is generated with random pair rapidity $Y$ and pair invariant mass $M$ according to the nuclear cross section. Then, pair momentum is calculated using generated values of $Y$ and $M$. Optionally, a non-zero transverse momentum of quasi-real photons and dilepton pairs can be taken into account as described in Section~\ref{sec:pt}. To assign momenta to each lepton in the final state, $z = \cos{\theta}$ is randomly generated according to the two-dimensional elementary cross section and the azimuthal angle $\phi$ is picked randomly in a range between $0$~and~$2\pi$.

Decays of the generated pair of tau leptons are simulated using either the Pythia~6 or the Pythia~8 event generator~\cite{pythia6,pythia8}; the user can choose between the two options. Parameters of generated leptons are passed to the Pythia decayer via an interface class \texttt{UpcPythia8Helper}~(or~\texttt{UpcPythia6Helper}). Finally the daughter particles are propagated to the particle stack. 

\subsection{Program structure}
\label{section:program_structure}

The simulation program, \generator, is coded in \CC. The target operational system is Linux; the program was successfully built and tested with the GCC~9.3 compiler. The generator code is organized as following:
\begin{itemize}
    \item the root folder contains the main source \texttt{main.cpp}, the build file \break \texttt{CMakeLists.txt}, the file with an example of input parameters \break \texttt{parameters.in}, the \texttt{README} file with a brief manual for installation and usage of the program and \texttt{include} and \texttt{src} folders containing files with all necessary calculation implementations.
    \item \texttt{\textbf{include}} contains \texttt{UpcGenerator.h} with a definition of the main class, \texttt{UpcCrossSection.h} with definition of a helper class for nuclear cross sections calculations, \texttt{UpcElemProcess.h} which is a base class for all elementary processes and where essential methods for elementary cross section calculations are defined, as well as \texttt{UpcPythia6Helper.h} and \texttt{UpcPythia8Helper.h} which implement simple interfaces for Pythia~6 and Pythia~8 respectively. The folder also contains auxiliary headers for logging in the subfolder \texttt{plog}, obtained from~\cite{plog_lib}.
    \item \texttt{src} contains source code files with implementations of methods for the program classes.
\end{itemize}

In the main source file, \texttt{main.cpp}, an instance of \texttt{UpcGenerator} class is created. The generator is initialized using parameters passed by a user via \texttt{parameters.in}. Initialization is done via a dedicated method \break \texttt{initGeneratorFromFile}, which reads all input parameters from file.

After initialization, a debug output level is specified. There are three levels of logging: a minimalistic logging with only essential information, such as information about the current calculation stage; a more verbose output, containing information about produced events and particles; the most verbose level of output, containing information about intermediate calculation steps for detailed debugging.

The simulation process is started via calling \texttt{generateEvents} method. Inside, the elementary cross section is calculated and stored into two-dimensional histogram via \texttt{fillCrossSectionZM} method, as well as nuclear cross section via \texttt{calcNucCrossSectionYM} method. Optionally, a transverse momentum cut is applied to leptons in final state. In the elementary cross section, a minimal value of $z$ is calculated for each $M$ bin to generate leptons with transverse momenta greater than a certain value, that can be specified by a user. After this step, the main simulation process starts, as it is described in Section~\ref{subsection:simulation_process}.

\subsection{Description of input parameters}
\label{subsection:input_parameters}

Upcgen can be controlled using parameters passed in \texttt{parameters.in} file. An example of the file with the default parameters can be found in the program repository (see the reference in Program Summary section). The existing input parameters are described below.
\begin{itemize}
    \item \textbf{NUCLEUS\_Z}, \textbf{NUCLEUS\_A}: nuclear charge and atomic number of incoming nuclei.
    \item \textbf{WS\_R}, \textbf{WS\_A}: spacial parameters of the Woods-Saxon distribution, corresponding to $R_A$ and $a$;
    \item \textbf{PROC\_ID}: number of a process to be generated. See a list of currently available processes below.
    \item \textbf{LEP\_A}: value of anomalous magnetic moment.
    \item \textbf{NEVENTS}: a number of events to be generated.
    \item \textbf{SQRTS}: energy of colliding system $\sqrt{s_{\rm NN}}$ at the center-of-mass frame.
    \item \textbf{DO\_PT\_CUT}: a flag that can be used to enable $p_{\rm T}$ cut for produced leptons.
    \item \textbf{PT\_MIN}: a minimal value of transverse momentum, which is used in $p_{\rm T}$ cut, if \textbf{DO\_PT\_CUT} is enabled.
    \item \textbf{ZMIN}, \textbf{ZMAX}, \textbf{MMIN}, \textbf{MMAX}, \textbf{YMIN}, \textbf{YMAX}: limits for \break $z = \cos{\theta}$, mass of a produced dilepton pair $M$ and pair rapidity $Y$ used in calculations of elementary and nuclear cross sections.
    \item \textbf{BINS\_Z}, \textbf{BINS\_M}, \textbf{BINS\_Y}: number of bins in each range mentioned in the previous item.
    \item \textbf{FLUX\_POINT}: a flag that can be used to switch between two ways of photon flux calculation: a flux from a point-like source and a flux obtained via integral~(see Section~\ref{section:theoretical_framework}).
    \item \textbf{NON\_ZERO\_GAM\_PT}: a flag that can be used to enable accounting for non-zero pair transverse momentum of produced dilepton pair.
    \item \textbf{USE\_POLARIZED\_CS}: a switch to account for photon polarization effects~(see Section~\ref{sec:polarization}).
    \item \textbf{PYTHIA\_VERSION}: a switch to choose between the versions of the Pythia event generator (either 6 or 8); by default, it is equal to $-1$, and Pythia is not used at all.
    \item \textbf{PYTHIA8\_FSR}: a flag that can be used to simulate final state radiation via Pythia~8.
    \item \textbf{PYTHIA8\_DECAYS}: a flag that can be used to enable lepton decays with Pythia~8 (decays are always done if the program is built with Pythia~6).
    \item \textbf{SEED}: a seed for the random number generator, which is used to draw random values of rapidity $Y$ and mass $M$ of a dilepton pair according to elementary cross section. The default value is $0$, which corresponds to a seed based on UNIX timestamp. A user-defined seed can be used to generate reproducible simulation results.
\end{itemize}

A list of available processes for the parameter \textbf{PROC\_ID} is the following:
\begin{itemize}
    \item \textbf{10}: Dielectron production $\gamma\gamma \to ee$.
    \item \textbf{11}: Dimuon production $\gamma\gamma \to \mu\mu$.
    \item \textbf{12}: Ditau production $\gamma\gamma \to \tau\tau$.
\end{itemize}
The list will be extended with further program developments in the future.

\subsection{Description of output data}
\label{subsection:output_data}

By default, parameters of generated particles in final state are stored in a ROOT file, in a tree. The tree contains following data:
\begin{itemize}
    \item \textbf{eventNumber}: a number of an event in which a particle was produced.
    \item \textbf{pdgCode}: a PDG code assigned according to Monte Carlo numbering scheme~\cite{ParticleDataGroup:2014cgo}.
    \item \textbf{particleID}: an identification number (ID) of a particle within an event.
    \item \textbf{statusID}: a status code according to the Pythia~8 definitions.
    \item \textbf{motherID}: an ID of a mother particle, which is equal to -1 in case of primary particles.
    \item $\boldsymbol{p_x}, \boldsymbol{p_y}, \boldsymbol{p_z}, \boldsymbol{E}$: components of four-momentum of a generated particle.
\end{itemize}
The output data is automatically compressed by ROOT, therefore such format can be used to efficiently store a great amount of events. It is also possible to store generated events using HepMC text-like format, which is widely used in high energy physics community~\cite{Buckley:2019xhk}.

\section{Description of test data}
\label{section:test_data}

The following input parameters were used to produce estimates for dimuon photoproduction in Pb--Pb UPCs shown in Fig.~\ref{fig:comparison_w_data}:
\begin{verbatim}
NUCLEUS_Z 82
NUCLEUS_A 208
WS_R 6.68
WS_A 0.447
SQRTS 5020
PROC_ID 11
LEP_A 0
NEVENTS 1000000
DO_PT_CUT 0
PT_MIN 0
ZMIN -1
ZMAX 1
MMIN 10
MMAX 50
YMIN -2.4
YMAX 2.4
BINS_Z 20000
BINS_M 2000
BINS_Y 200
FLUX_POINT 0
NON_ZERO_GAM_PT 1
USE_POLARIZED_CS 1
PYTHIA_VERSION 8
PYTHIA8_FSR 1
PYTHIA8_DECAYS 0
SEED 0
\end{verbatim}
In this simulation, we produced a total number of 500\,000\,000 events in multiple program runs with an additional dimuon mass cut $m_{\mu\mu} > 10~{\rm GeV}/c^2$ in order to reach reasonable level of statistical uncertainties in the mass region of interest. The generated events were selected using the kinematic selections discussed in Section~\ref{section:theoretical_framework} using a simple \CC/ROOT program to extract fiducial cross sections $\mathrm{d}\sigma/\mathrm{d}m_{\mu\mu}$ and $\mathrm{d}\sigma/\mathrm{d}\alpha$. 
\section{Summary and outlook}
\label{section:summary}

Upcgen is an event generator dedicated to photoproduction of dilepton pairs in ultra-peripheral collisions of heavy ions. An accurate treatment of the photon flux at small impact parameters allows us to improve the description of dimuon production cross sections measured by ATLAS~\cite{DimuonATLAS} at large rapidities of the dimuon pair. An accurate calculation of the dilepton pair production in UPCs becomes especially important in view of future precision measurements at the LHC in Run 3 and 4.

The \generator~generator allows a user to compute ditau cross sections for arbitrary values of the tau anomalous magnetic moment. This feature is useful for sensitivity studies of future $a_{\tau}$ measurements via ditau production in UPCs.

Although the calculation of cross sections in the generator takes into account various subtle aspects, it does not include higher-order QED effects, such as unitary and Coulomb corrections~\cite{Hencken:2006ir}. In particular, the Coulomb corrections, related to possible photon exchanges between final state leptons and heavy ions, is a longly debated subject. Hencken~{\it et al.}~\cite{Hencken:2006ir} argued that Coulomb corrections are strongly suppressed due to cancellations of diagrams involving positively and negatively charged leptons, however recent studies suggest that Coulomb corrections might be significant~\cite{Zha:2021jhf}. Besides, the theoretical model used in \generator~only accounts for coherent emission of photons from nuclei and omits effects arising from dissociative processes, in which one photon comes off from charged constituents of a nucleus~\cite{Baur:1998ay}. These processes will be considered in future versions of \generator.

\section{Acknowledgements}
\label{section:acknowledgements}
We are grateful to  Mateusz Dyndal for his suggestion to include FSR effects and to Lucian Harland–Lang, Valeri Khoze and Mikhail Ryskin for drawing our attention to photon polarization effects and SuperChic results.

This work was supported by the Russian Foundation for Basic Research according to the project no.~21-52-14006 and the Austrian Science Fund according to the project no.~I~5277-N.


\bibliographystyle{elsarticle-num}
\bibliography{bibliography}

\end{document}